\documentclass[%
 reprint,
aps,
showpacs,
 amsmath,
prl,
]{revtex4-1}

\newcommand{\be}{\begin{equation}}
\newcommand{\ee}{\end{equation}}

\begin{document}

\preprint{APS/123-QED}

\title{Thermodynamic derivation and use of a nonequilibrium canonical ensemble}

\author{Maarten H. P. Ambaum}
\affiliation{%
 Department of Meteorology,
 University of Reading, U.K.
}%

\date{\today}

\begin{abstract}
\noindent A thermodynamic expression for the analog of the canonical ensemble for nonequilibrium systems is described based on a purely information theoretical interpretation of entropy.  As an application, it is shown that this nonequilibrium canonical distribution implies some important results from nonequilibrium thermodynamics, specifically, the fluctuation theorem and the Jarzynski-equality.

\end{abstract}

\maketitle


\noindent In this letter we demonstrate that the information-theoretical definition of entropy implies some important results in non-equilibrium thermodynamics, such as the fluctuation theorem and the Jarzynski-equation.  The central tenet is that for two states $A$ and $B$ of a system, defined by two sets of macroscopic parameters, the ratio of the probabilities $p_B/p_A$ for the system to be in either state is
\be
p_B/p_A = \exp (\Delta_{AB}S/k_B),
\ee
with $\Delta_{AB}S$ difference in entropy between the states $B$ and $A$.  This is essentially the Boltzmann definition of entropy, combined with the observation that in the absence of any other information, each state is assumed to have equal probability.  The latter assumption is in an information-theoretical setting equivalent to the \textsl{principle of indifference}: the absence of any distinguishing information is equivalent to equal prior (prior to obtaining additional information) probabilities \cite{prior}.

Following Boltzmann, we define the entropy $S$ as the logarithm of the number of states accessible to a system under given macroscopic constraints.  For an isolated system, the entropy is related to the size $\Phi$ of the accessible phase space,
\be 
S = k_B\ln \Phi.
\ee
For a classical system, the phase space size $\Phi$ is the hyper-area of the energy shell, and it defines the usual microcanonical ensemble.  The hyper-area $\Phi$ is non-dimensionalised such that $\Phi(U)\,\text{d}U$ is proportional to the number of states between energies $U$ and $U+\text{d}U$.  We will not consider other required factors which make the argument of the logarithm non-dimensional; these contribute an additive entropy constant which will not be of interest to us here.  Note also that the microcanonical ensemble does not include a notion of equilibrium: the system is assumed to be insulated so it cannot equilibrate with an external system.  It just moves around on the energy shell and the assumed ergodicity implies that all states, however improbable from a macoscopic point of view, are members of the ensemble.  Of course, the number of unusual states (say, with non-uniform macroscopic density) is much lower than the number of regular states (say, with uniform macroscopic density) for macroscopic systems.  Only for small systems, the distinction becomes important but it does not invalidate the above formal definition of entropy.
The above definition of entropy also ensures that entropy is an extensive property such that for two independent systems considered together the total entropy is the sum of the individual entropies, $S=S_1+S_2$.  The Boltzmann constant $k_B$ ensures dimensional compatibility with the classical thermodynamic entropy when the usual equilibrium assumptions are made \cite{jaynes1,*gibbsvboltzmann}.

The hyper-area of the energy shell, and thus the entropy, can be a function of several variables which are set as external constraints, such as the total energy $U$, system volume, $V$, or particle number $N$.  For the canonical ensemble we consider a system that can exchange energy with some reservoir.  We consider here only a theoretical canonical ensemble in that we consider the coupling between the two systems to be weak.

First, we need to define what a reservoir is.  Following equilibrium thermodynamics, we formally define an `inverse temperature' $\beta=(k_BT)^{-1}$ as 
\be
\beta = k_B^{-1}\partial S/\partial U = \Phi^{-1}\partial \Phi/\partial U.       \label{invt}
\ee
We make no claim about the equality of $\beta$ and the classical equilibrium inverse temperature; $\beta$ is the expansivity of phase space with energy and as such can be defined for any system, whether it is in thermodynamic equilibrium or not.  When an isolated system is prepared far from equilibrium (for example, when it has a local equilibrium temperature which varies over the system) then $\beta$ is still uniquely defined for the system as a non-local property of the energy shell that the system resides on.  Because both energy and entropy in the weak coupling limit are extensive quantities, $\beta$ must be an intensive quantity.

Now consider a large isolated system $R$ with total (internal) energy $U_R$.  Let this system receive energy $U'$ from the environment.  By expanding $S$ in powers of $U$, we can then write the entropy of this large system as
\be
S_R(U_R+U') = S_R(U_R)+k_BU' \left(\beta + \frac{1}{2}U'\frac{\partial\beta}{\partial U}+\mathcal{O}(U'^2)\right).
\ee
We see that for finite $U',$ $(\partial\beta/{\partial U})^{-1}$ has to be an extensive quantity.  But that means that for a very large system $\partial\beta/{\partial U} = \mathcal{O}(N^{-1}),$ where $N$ is a measure of the size of the system (such as particle number).  For a classical thermodynamic system $\partial\beta/{\partial U} = -k_B\beta^2/C_V$ with $C_V$ the heat capacity at constant volume.  We conclude that for a very large system ($N \rightarrow\infty$), the entropy equals
\be
S_R(U_R+U') = S_R(U_R) + k_B\beta U'          \label{entropy}
\ee
for all relevant, finite energy exchanges $U'$.  This expression for the entropy defines a reservoir.  The size of the energy shell accessible to the reservoir is, for all relevant energy exchanges $U'$, exactly proportional to $\exp(\beta U'),$ with $\beta$ an intensive and constant property of the reservoir.  We do not require the reservoir to be in thermodynamic equilibrium.  A change of energy in the reservoir pushes the reservoir to a different energy shell; the functional dependence of the size of the energy shell with energy defines the `inverse temperature' $\beta$, as in Eq.~\ref{invt}.  However, it is not assured that a small and fast thermometer would measure an inverse temperature equal to $\beta$ at some point in the reservoir; only if the reservoir is allowed to equilibrate, its inverse temperature is everywhere equal to $\beta$.  Of course, this is precisely how the temperature of a reservoir is determined in practice.

Now suppose a system of interest has energy $U_0$.  We then allow it to exchange heat $U$ with a reservoir.  If the system has energy $U_0+U$, the reservoir must have given up energy $U$.  We can write the hyper-area of the energy shell of the system $\Phi_0$ as a function of $U$.  The total entropy of the system plus reservoir $R$ can then be written as 
\be
S = S_0(U) + S_R(U_R)-k_B\beta U,
\ee
with $S_0 = k_B\ln\Phi_0$.
The number of states at each level of exchange energy therefore is proportional to 
\be
\Phi (U) \propto \Phi_0(U)\,\exp(-\beta U),    \label{density}
\ee
where we omitted proportionality constants related to the additive entropy constants.  Nowhere we assume that the system is in equilibrium with the reservoir.  This means that $\Phi(U)$ is the relevant measure to construct an ensemble average for the system, even for far-from-equilibrium systems.  Even the reservoir can be out of equilibrium, as discussed above. We have also made no reference to the size of the system of interest, as long as it is much smaller than the reservoir.  However, in contrast to systems in thermodynamic equilibrium, there is no guarantee that the extensive variables, such as $U$, $V$, or $N$ define the state of the system in any reproducible sense.  To fully define an out-of-equilibrium system we need to introduce order parameters that can describe the non-equilibrium aspects of the system.

The above density is an integrated version of the usual canonical distribution.  The size of the energy shell of the system of interest, $\Phi_0$, can be written as an integral over states $\Gamma$ such that
\be
\Phi_0(U) = \int_{H_0(\Gamma)=U} \text{d}\Gamma,
\ee
with $H_0$ the Hamiltonian of the system of interest .
With this definition, the density in Eq.~\ref{density} reduces to the usual canonical distribution  $\exp(-\beta H_0(\Gamma))$ for states $\Gamma$.  We will not make further use of this microscopic version of the density.

The canonical density in Eq.~\ref{density} can be expanded by parametrizing each energy shell with some continuous coordinate $v,$ so that every part of phase space has coordinates $(U,v)$.   At each value of $v$ the differential $\phi(U,v)\,\text{dU}\,\text{d}v$ is proportional to the number of states between coordinate values $U$ and $U+\text{d}U$, and $v$ and $v+\text{d}v$, and it is normalised such that 
\be
\int \phi(U,v)\,\text{d}v = \Phi_0(U).
\ee
The parametrisation is arbitrary at this point and can be chosen such as to divide the phase space in as fine a structure as desired for a given application.  We can define an entropy $S_0(U,v)$ again as the logarithm of the number of available states for the system of interest corresponding to point $(U,v)$,
\be
S_0(U,v) = k_B\ln \phi(U,v).
\ee

Now consider a process that occurs on the energy shell $U$ where some variable changes from $A\rightarrow B$ .  On the parametrized energy shell this corresponds to a coordinate shift from $v(A) \rightarrow v(B)$.  The number of corresponding states changes from $\phi(U,v(A))\rightarrow \phi(U,v(B))$.  We can use detailed balance to express the ratio of the probability of making this transition to the probability of making the reverse transition as the ratio of the number of states at $(U,v(A))$ to the number of states at $(U,v(B))$:
\be
\frac{p_{A\rightarrow B}}{p_{B\rightarrow A}} = \frac{\phi(U,v(B))}{\phi(U,v(A))} = \exp (\Delta_{AB}S/k_B),      \label{FT0}
\ee
where $\Delta_{AB}S/k_B = S_0(U,v(B))- S_0(U,v(A))$.  The Liouville theorem implies that the local phase space volume is conserved when a system moves on the energy shell, so that $\Delta_{AB}S$ has to vanish for every realisable transformation.  However, this is not the case anymore if the system moves between energy shells.  If, in addition, during the process $A\rightarrow B$ the energy of the system of interest changes from $U_A\rightarrow U_B$ through exchange with the reservoir, then the above ratio of probabilities can still be expressed as $\exp (\Delta_{AB}S/k_B)$ but now with
\be
\Delta_{AB}S = S_0(U_B,v(B))- S_0(U_A,v(A))-k_B\beta (U_B-U_A).
\ee
We can always write the entropy change of the system of interest as the sum of the entropy change due to heat exchange with the reservoir and an irreversible entropy change associated with uncompensated heat \cite{kondepudi,*ambaum}, viz. $S_0(U_B,v(B))- S_0(U_A,v(A)) = k_B\beta (U_B-U_A) + \Delta_i S_0$.  We thus conclude that $\Delta_{AB}S = \Delta_iS_0$, that is, the relevant entropy change in Eq.~\ref{FT0} equals the irreversible entropy change of the system of interest.  So for processes that occur either on or across energy shells, we have
\be
\frac{p_{A\rightarrow B}}{p_{B\rightarrow A}} = \exp (\Delta_iS_0/k_B),
\label{FTAB} 
\ee
with $\Delta_iS_0$ the irreversible entropy change of the system in a process $A\rightarrow B.$    The right-hand-side of this equation is only dependent on the irreversible entropy change $\Delta_iS_0$ between the two states of the system of interest.  So this equation must be true for any pair of states $(A,B)$ that are related by the same irreversible entropy change.  We thus arrive at the \textsl{fluctuation theorem} \cite{evans1,*evans},
\be
\frac{p(\Delta_iS)}{p(-\Delta_iS)} = \exp (\Delta_iS/k_B),      \label{FT}
\ee
with $p(\Delta_iS)$ the probability the system of interest makes a transition with irreversible entropy change of $\Delta_iS$ and $p(-\Delta_iS)$ the probability for the opposite change.

The fluctuation theorem applies to spontaneous processes that occur in thermostatted but otherwise isolated systems.  We next consider processes that occur when we modify the system of interest by changing some external macroscopic parameters.  The entropy of the energy shell $U$ is then also a function of some parameter $\lambda$, viz., $S=S_\lambda(U,v)$.  Without loss of generality we set $\lambda=0$ at $A$ and $\lambda=1$ at $B$.  In this case the irreversible entropy change in, Eq.~\ref{FTAB}, is
\be
\Delta_iS/k_B =  S_1(U_B,v(B))- S_0(U_A,v(A))-k_B\beta (U_B-U_A).
\ee
Apart from this, there is no change in the considerations leading to the fluctuation theorem.  By definition, thermostatted systems that receive work $W_{AB}$ from their environment have an irreversible entropy change equal to
\be
\Delta_iS/k_B = \beta (W_{AB}-\Delta_{AB} F),       \label{workfunction}
\ee
with $\Delta_{AB} F$ the change in free energy going from $A$ to $B$.
Recognising that the right-hand-side is again only a function of the difference between the two states, we arrive at the \textsl{Crooks fluctuation theorem} \cite{crooks},
\be
\frac{p_{01}(W)}{p_{10}(-W)} = \exp (\beta (W-\Delta_{01} F)),       \label{crooks}
\ee
with $p_{01}(W)$ the probability that the system absorbs work $W$ when $\lambda$ changes from 0 to 1, and $p_{10}(-W)$ the probability that the system performs work $W$ when $\lambda$ changes in reverse from 1 to 0.
Because the transition probabilities can be normalised with respect to the exchanged work, it is straightforward to use this equation to show that the expectation value of $\exp (-\beta (W-\Delta_{01} F))$ equals unity, or equivalently,
\be
\langle \exp (-\beta W) \rangle =  \exp (-\beta \Delta_{01} F).
\label{jarzynski}
\ee 
This is the Jarzynski equation \cite{jarzynski}.

The consistency of the above argument is strengthened by the following independent route to calculate free energy changes. 
The phase space measure $\Phi(U)$ can be normalised with the partition function $Z_\lambda$,
\be 
Z_\lambda = \int \Phi_\lambda (U)\,\exp(-\beta U)\,\text{d}U.
\ee
where $\Phi_\lambda (U)$ is proportional to the number of accessible states of the isolated the system of interest when the external parameter is set to $\lambda$.
The equilibrium free energy for the thermostated system is
\be
F_\lambda = -\beta^{-1}\ln Z_\lambda.                    \label{free}
\ee
Next we consider what happens to the equilibrium free energy of the system when we vary $\lambda$ from 0 to 1. The partition function at $\lambda=1$ satisfies
\begin{align}
Z_1 &= \int \Phi_1(U) \,\exp(-\beta U)\,\text{d}U \\
&= \int \Phi_0(U)\,\exp(\Delta S/k_B) \,\exp(-\beta U)\,\text{d}U\\
&= Z_0 \, \langle\exp(\Delta S/k_B)\rangle
\end{align}
where $\langle .\rangle$ denotes an ensemble average over the initial ensemble, and $\Delta S = k_B\ln(\Phi_1(U)/\Phi_0(U)$.  As before, the entropy change can be written as the sum of the entropy change due to heat exchange with the reservoir and the irreversible entropy change due to uncompensated heat.  Because the system plus the reservoir are thermally insulated, any heat given to the reservoir must be compensated by work performed by the external parameter change.  The entropy change, above can therefore be written as $
 \langle\exp(\Delta S/k_B)\rangle =  \langle\exp(\Delta_iS/k_B-\beta W)\rangle
$  so that we find
\begin{align}
Z_1/Z_0 =  \langle\exp(\Delta_iS/k_B-\beta W)\rangle.
\end{align}
Because Eq.~\ref{workfunction} is true for any microscopic realisation of the process, we find that the right-hand-side of the above equation is the same for every realisation  and it is equal to $\exp(-\beta\Delta F)$.  This is consistent with the equilibrium expression for the free energy, Eq.~\ref{free}, from which folllows that $\exp(-\beta\Delta F) = Z_1/Z_0.$  The above equation is only apparently in contradiction to the Jarzynski equation, Eq.~\ref{jarzynski}.  To arrive at the Jarzynski equation  we recognise that Eq.~\ref{workfunction} implies that $\langle\exp(\beta(\Delta F- W))\rangle = \langle\exp(-\Delta_iS/k_B)\rangle =1,$ where the last equality follows from integrating the fluctuation theorem over all values of $\Delta_iS$.

\bibliography{noneqens,noneqensNotes}

\begin{thebibliography}{1}%
\makeatletter
\providecommand \@ifxundefined [1]{%
 \ifx #1\undefined \expandafter \@firstoftwo
 \else \expandafter \@secondoftwo
\fi
}%
\providecommand \@ifnum [1]{%
 \ifnum #1\expandafter \@firstoftwo
 \else \expandafter \@secondoftwo
\fi
}%
\providecommand \enquote [1]{``#1''}%
\providecommand \bibnamefont  [1]{#1}%
\providecommand \bibfnamefont [1]{#1}%
\providecommand \citenamefont [1]{#1}%
\providecommand\href[0]{\@sanitize\@href}%
\providecommand\@href[1]{\endgroup\@@startlink{#1}\endgroup\@@href}%
\providecommand\@@href[1]{#1\@@endlink}%
\providecommand \@sanitize [0]{\begingroup\catcode`\&12\catcode`\#12\relax}%
\@ifxundefined \pdfoutput {\@firstoftwo}{%
 \@ifnum{\z@=\pdfoutput}{\@firstoftwo}{\@secondoftwo}%
}{%
 \providecommand\@@startlink[1]{\leavevmode\special{html:<a href="#1">}}%
 \providecommand\@@endlink[0]{\special{html:</a>}}%
}{%
 \providecommand\@@startlink[1]{%
  \leavevmode
  \pdfstartlink
   attr{/Border[0 0 1 ]/H/I/C[0 1 1]}%
   user{/Subtype/Link/A<</Type/Action/S/URI/URI(#1)>>}%
  \relax
 }%
 \providecommand\@@endlink[0]{\pdfendlink}%
}%
\providecommand \url  [0]{\begingroup\@sanitize \@url }%
\providecommand \@url [1]{\endgroup\@href {#1}{\urlprefix}}%
\providecommand \urlprefix [0]{URL }%
\providecommand \Eprint[0]{\href }%
\@ifxundefined \urlstyle {%
  \providecommand \doi [1]{doi:\discretionary{}{}{}#1}%
}{%
  \providecommand \doi [0]{doi:\discretionary{}{}{}\begingroup
  \urlstyle{rm}\Url }%
}%
\providecommand \doibase [0]{http://dx.doi.org/}%
\providecommand \Doi[1]{\href{\doibase#1}}%
\providecommand \bibAnnote [3]{%
  \BibitemShut{#1}%
  \begin{quotation}\noindent
    \textsc{Key:}\ #2\\\textsc{Annotation:}\ #3%
  \end{quotation}%
}%
\providecommand \bibAnnoteFile [2]{%
  \IfFileExists{#2}{\bibAnnote {#1} {#2} {\input{#2}}}{}%
}%
\providecommand \typeout [0]{\immediate \write \m@ne }%
\providecommand \selectlanguage [0]{\@gobble}%
\providecommand \bibinfo [0]{\@secondoftwo}%
\providecommand \bibfield [0]{\@secondoftwo}%
\providecommand \translation [1]{[#1]}%
\providecommand \BibitemOpen[0]{}%
\providecommand \bibitemStop [0]{}%
\providecommand \bibitemNoStop [0]{.\EOS\space}%
\providecommand \EOS [0]{\spacefactor3000\relax}%
\providecommand \BibitemShut [1]{\csname bibitem#1\endcsname}%
\bibitem{prior}%
  \BibitemOpen
  \bibfield{author}{%
  \bibinfo {author} {\bibfnamefont{E.~T.}\ \bibnamefont{Jaynes}},\ }%
  \bibfield{journal}{%
  \bibinfo {journal} {IEEE Transactions On Systems Science and Cybernetics}\ }%
  \textbf{\bibinfo {volume} {4}},\ \bibinfo {pages} {227} (\bibinfo {year}
  {1968})%
  \bibAnnoteFile{NoStop}{prior}%
\bibitem{jaynes1}%
  \BibitemOpen
  \bibfield{author}{%
  \bibinfo {author} {\bibfnamefont{E.~T.}\ \bibnamefont{Jaynes}},\ }%
  \bibfield{journal}{%
  \bibinfo {journal} {The Physical Review}\ }%
  \textbf{\bibinfo {volume} {4}},\ \bibinfo {pages} {620} (\bibinfo {year}
  {1957})%
  \bibAnnoteFile{NoStop}{jaynes1}%
\bibitem{gibbsvboltzmann}%
  \BibitemOpen
  \bibfield{author}{%
  \bibinfo {author} {\bibfnamefont{E.~T.}\ \bibnamefont{Jaynes}},\ }%
  \bibfield{journal}{%
  \bibinfo {journal} {American Journal of Physics}\ }%
  \textbf{\bibinfo {volume} {5}},\ \bibinfo {pages} {391} (\bibinfo {year}
  {1965})%
  \bibAnnoteFile{NoStop}{gibbsvboltzmann}%
\bibitem{kondepudi}%
  \BibitemOpen
  \bibfield{author}{%
  \bibinfo {author} {\bibfnamefont{D.}~\bibnamefont{Kondepudi}}\ and\ \bibinfo
  {author} {\bibfnamefont{I.}~\bibnamefont{Prigogine}},\ }%
  \emph{\bibinfo {title} {Modern thermodynamics}}\ (\bibinfo {publisher} {J.
  Wiley \& Sons},\ \bibinfo {address} {Chichester},\ \bibinfo {year} {1998})%
  \bibAnnoteFile{NoStop}{kondepudi}%
\bibitem{ambaum}%
  \BibitemOpen
  \bibfield{author}{%
  \bibinfo {author} {\bibfnamefont{M.~H.~P.}\ \bibnamefont{Ambaum}},\ }%
  \emph{\bibinfo {title} {Thermal physics of the atmosphere}}\ (\bibinfo
  {publisher} {Wiley--Blackwell},\ \bibinfo {address} {Chichester},\ \bibinfo
  {year} {2010})%
  \bibAnnoteFile{NoStop}{ambaum}%
\bibitem{evans1}%
  \BibitemOpen
  \bibfield{author}{%
  \bibinfo {author} {\bibfnamefont{D.~J.}\ \bibnamefont{Evans}}, \bibinfo
  {author} {\bibfnamefont{E.~G.~D.}\ \bibnamefont{Cohen}},\ and\ \bibinfo
  {author} {\bibfnamefont{G.~P.}\ \bibnamefont{Morriss}},\ }%
  \bibfield{journal}{%
  \Doi{10.1103/PhysRevLett.71.2401}{\bibinfo {journal} {Phys. Rev. Lett.}}\ }%
  \textbf{\bibinfo {volume} {71}},\ \bibinfo {pages} {2401} (\bibinfo {month}
  {Oct}\ \bibinfo {year} {1993})%
  \bibAnnoteFile{NoStop}{evans1}%
\bibitem{evans}%
  \BibitemOpen
  \bibfield{author}{%
  \bibinfo {author} {\bibfnamefont{D.~J.}\ \bibnamefont{Evans}}\ and\ \bibinfo
  {author} {\bibfnamefont{D.~J.}\ \bibnamefont{Searles}},\ }%
  \bibfield{journal}{%
  \Doi{10.1080/00018730210155133}{\bibinfo {journal} {Advances in Physics}}\ }%
  \textbf{\bibinfo {volume} {51}},\ \bibinfo {pages} {1529} (\bibinfo {year}
  {2002})%
  \bibAnnoteFile{NoStop}{evans}%
\bibitem{crooks}%
  \BibitemOpen
  \bibfield{author}{%
  \bibinfo {author} {\bibfnamefont{G.~E.}\ \bibnamefont{Crooks}},\ }%
  \bibfield{journal}{%
  \Doi{10.1103/PhysRevE.60.2721}{\bibinfo {journal} {Phys. Rev. E}}\ }%
  \textbf{\bibinfo {volume} {60}},\ \bibinfo {pages} {2721} (\bibinfo {month}
  {Sep}\ \bibinfo {year} {1999})%
  \bibAnnoteFile{NoStop}{crooks}%
\bibitem{jarzynski}%
  \BibitemOpen
  \bibfield{author}{%
  \bibinfo {author} {\bibfnamefont{C.}~\bibnamefont{Jarzynski}},\ }%
  \bibfield{journal}{%
  \bibinfo {journal} {Physical Review Letters}\ }%
  \textbf{\bibinfo {volume} {78}},\ \bibinfo {pages} {2690} (\bibinfo {year}
  {1997})%
  \bibAnnoteFile{NoStop}{jarzynski}%
\end{thebibliography}%

\end{document}